# Single-Sideband Time-Modulated Phased Array With 2-bit Phased Shifters

Yanchang Gao, Gang Ni, Kun Wang, Yiqing Liu, Chong He, Ronghong Jin, Xianling Liang
Department of Electronic Engineering
Shanghai Jiao Tong University
Shanghai 200240, China
hechong@sjtu.edu.cn

*Abstract*- **A novel single-sideband (SSB) time-modulated technique with 2-bit phase shifters is proposed. The time-modulated module is implemented by adding periodic phase modulation to 2-bit phase shifters, which is simpler without performance loss compared to existing SSB time-modulated method. During one modulation period, four phase states (0, $\pi/2$, $\pi$, $3\pi/2$) of 2-bit phase shifters are switched in sequence. After the modulation, the SSB time modulation is realized and the main power is distributed to the first harmonic component. The feasibility of the proposed method is verified by experiments. The undesired harmonics are efficiently suppressed. Meanwhile, $\pm 40°$ beam scanning range are realized through the proposed module.**

## I. INTRODUCTION

Conventional phased array system includes plenty of phase shifters and attenuators. For high accuracy and low sidelobe, 6-bit or more digital phase shifters and variable attenuators are usually exploited. The use of large number of phase shifters and attenuators makes the control module complicated, which further increases the cost and complexity of the phased array. Time-modulated arrays have the advantages of low cost and high accuracy of amplitude and phase control. However, their efficiency and flexibility need to be improved [1].

The efficiency of time-modulated array is related to feeding network efficiency and harmonic efficiency, and there have been many related studies. To improve feeding network efficiency, the reconfigurable power divider is applied to time-modulated array [2]. In order to improve harmonic efficiency, 1-bit phase shifters are utilized to suppress the fundamental and even harmonic components in [3]. And in [4], an in-phase/quadrature (I/Q) complex modulation technique was proposed to realize single-sideband time-modulated phase only weighting. In [5], an SSB time-modulated module with multiple branches was proposed to suppress most of the unwanted harmonic components. In addition, the reconfigurable power divider was used in the I/Q modulator [6], which can improve feeding network efficiency and harmonic efficiency simultaneously.

In this conference paper, we propose a novel SSB time-modulated method with 2-bit phase shifters. During one modulation period, four states (0, $\pi/2$, $\pi$, $3\pi/2$) of the 2-bit phase shifters are switched in sequence to concentrate the main power into the first harmonic, while useless harmonics are suppressed. The proposed method has a simpler structure compared to existing SSB time-modulated methods.

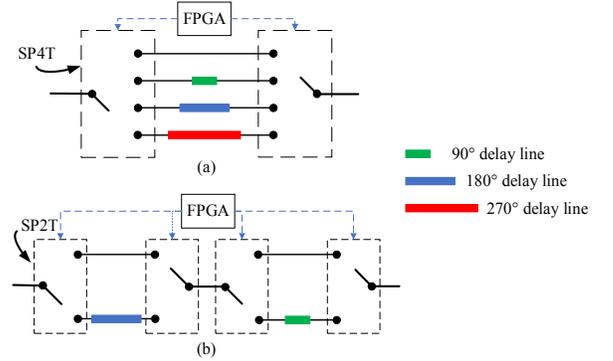

Figure 1. Configuration of the proposed module based on: (a) SP4T and (b) SP2T switches.

The remainder of this manuscript is organized as follows. Section II provides the mathematical theory of the proposed module. In Section III, a L-band 8-element time-modulated array with the proposed module is fabricated and tested to verify its effectiveness. Finally, some conclusions are drawn in Section IV.

## II. MATHEMATICAL FORMULATION

Fig. 1 shows the configuration of the proposed module. The RF switches are modulated periodically with the function

$$U(t) = \begin{cases} 1, & t_{1n}+mT_p < t < t_{1n}+\tau+mT_p \\ e^{j\pi/2}, & t_{2n}+mT_p < t < t_{2n}+\tau+mT_p \\ -1, & t_{3n}+mT_p < t < t_{3n}+\tau+mT_p \\ e^{j3\pi/2}, & t_{4n}+mT_p < t < t_{4n}+\tau+mT_p \end{cases}, \quad (1)$$

where $T_p$ is the modulation period ($f_p=1/T_p$), $t_{1n}$, $t_{2n}$, $t_{3n}$, $t_{4n}$ are the switch-ON time instants, $\tau$ is the duration of switch-ON time ($\tau \leqslant 0.25T_p$).

$U(t)$ can be decomposed by the Fourier series in frequency domain as

$$U(t) = \sum_{h=-\infty}^{\infty} A_{hn} e^{j2\pi h f_p t}. \quad (2)$$

Assume that

$$t_{3n} - t_{1n} = t_{4n} - t_{2n} = 0.5T_p, \quad (3)$$

$$t_{1n} = t_{2n} + 1/4 T_p. \quad (4)$$

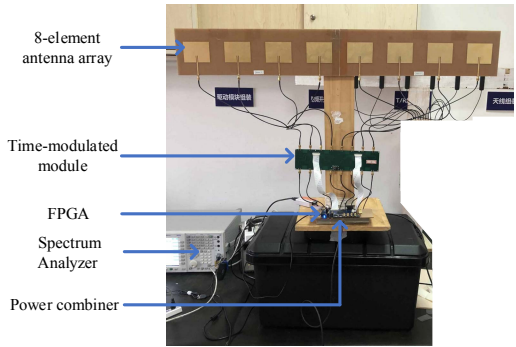

Figure 2. Photograph of the 8-element TMA with the proposed module.

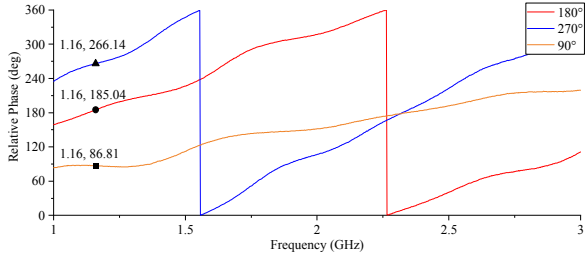

Figure 3. Relative phases of three delay lines compared to the reference.

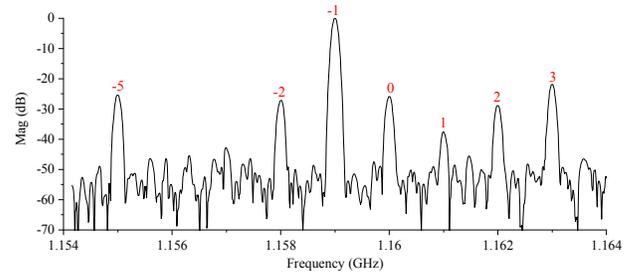

Figure 4. Measured power spectrum after the periodical phase modulation.

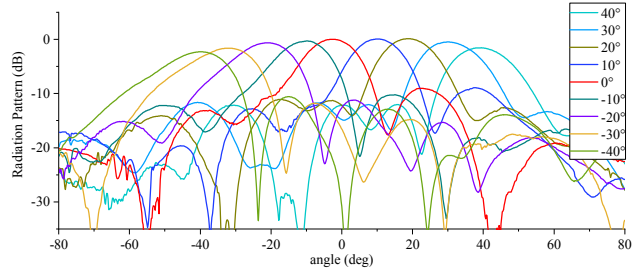

Figure 5. Scanning performance of the 8-element TMA.

Based on the assumption, the output is calculated as

$$s_o(t) = \begin{cases} 0, & h=0 \\ \sum_{h=-\infty}^{\infty} F e^{-j\pi h f_p(2t_{1n}+\tau)} \left( e^{j\pi \frac{1+h}{2}} + 1 \right) e^{j2\pi f_c t}, & h \neq 0 \end{cases} \quad (5)$$

where $f_c$ is the carrier frequency, and $F$ is written as

$$F = \frac{2}{h\pi} \sin(h\pi\tau f_p) \sin\left(\frac{h\pi}{2}\right). \quad (6)$$

As can be seen from (5) and (6), $s_o(t) = 0$ at the harmonics of $h = 2k \cup 4k+1$ ($k \in \mathbb{Z}$). Hence, the fundamental, $(2k)^{th}$ and $(4k-1)^{th}$ harmonic components are suppressed through this module. The output signal contains the harmonic components with the order -1, 3, -5, 7 and so on.

The main power of the RF signal is transferred the 1st harmonic component. If $\tau = 0.25 T_p$, the ideal loss is

$$IL = 20\lg(2\sqrt{2}/\pi) = -0.91 dB. \quad (7)$$

### III. EXPERIMENTAL RESULTS

To prove the feasibility of the proposed method, a set of experiments were completed by an 8-element time-modulated array. The carrier frequency is 1.16 GHz and the modulation frequency is 1 MHz. The relative phases of those delay lines are provided in Fig. 3 and the max phase error is 5°.

First, the power spectrum of the signal after the periodic phase modulation is measured, as shown in Fig. 4. Due to the phase shift errors, the fundamental and several harmonics (−2nd and 2nd) are not eliminated completely, but the levels of those components are less than −25 dB. Additionally, the main power is transferred to the −1st harmonic component.

Then, the pattern of the −1st harmonic component is examined by an 8-element array. The elements spacing is half wavelength and the center frequency of the antenna array is 1.16 GHz. The received signal is modulated by the module and combined by the power combiner, and then input to the spectrum analyzer. By adjusting the modulation signal to change the phase of each element, beam scanning angles of ±40 degrees are obtained with stepping 10 degrees and the power patterns of the −1st harmonic component are plotted in Fig. 5.

### IV. CONCLUSION

In this manuscript, a novel SSB time-modulated method utilizing 2-bit phase shifters is proposed to suppress undesired harmonic components. Compared to the existing SSB time-modulated approaches, the proposed technique is simpler in structure. From the theoretical and experimental results, less power loss is caused by the periodic modulation and accurate phase control is implemented. The further research locates the combined amplitude-phase control with the proposed module.